\documentclass{article}
\usepackage{spconf,amsmath,mathrsfs,amssymb,graphicx}
\usepackage{booktabs,bm,float,color,stfloats}
\usepackage{algorithm}
\usepackage{algpseudocode}
\usepackage{url}
\makeatletter
\newcommand{\figcaption}{\def@captype{figure}\caption}
\newcommand{\tabcaption}{\def@captype{table}\caption}
\makeatother
\title{Deep graph convolution neural network with non-negative matrix factorization for community discovery}
\name{Shuliang Xu$^{\dagger}$, Shenglan Liu$^{\ddagger}$ and Lin Feng$^{\ddagger,*}$\thanks{*Corresponding author.}}
\address{$\dagger$ Faculty of Electronic Information and Electrical Engineering, Dalian University of Technology\\
$\ddagger$ School of Innovation and Entrepreneurship, Dalian University of Technology\\
slx\_cs@mail.dlut.edu.cn, \{liusl, fenglin\}@dlut.edu.cn}

\begin{document}
%\ninept
%
\maketitle
\begin{abstract}
Community discovery is an important task for graph mining. Owing to the nonstructure, the high dimensionality and the sparsity of graph data, it is not easy to obtain an appropriate community partition. In this paper, a deep graph convolution neural network with non-negative matrix factorization (DGCN-NMF) is proposed for community discovery. DGCN-NMF employs multiple graph convolution layers to obtain the low dimensional embedding. In each layer, the last output and the non-negative matrix factorization's results of the previous outputs are inputted into the current layer. The community partition of the inputted graph can be outputted by an end-to-end approach. The proposed algorithm and the comparison algorithms are conducted on the experimental data sets. The experimental results show that the proposed algorithm outperforms the comparison algorithms on the experimental data sets. The experimental results demonstrate that DGCN-NMF is an effective algorithm for community discovery.
\end{abstract}
\section{Introduction}
Community discovery is an important approach for graph mining. It partitions the nodes of a graph into multiple communities. The nodes in the same community are densely connected and the nodes in different communities are sparsely connected \cite{coscia2019discovering,ni2019local}. Community discovery has been applied to many applications such as social network analysis \cite{jiang2019efficient}, recommended system \cite{liu2016trajectory} and geographic information system \cite{rozenshtein2019mining}, etc. Adjacency matrix is an important representation of a graph. However, due to the non-structure, the high dimension and the sparsity of graph data, it brings great challenges for community discovery algorithms \cite{LiuDeep2020}. How to effectively mine community structure in graph data has attracted much attention from researchers. Up to now, many community discovery algorithms have been proposed. They can be divided into three categories: modularity optimization, matrix factorization and graph neural network.

Modularity optimization is a heuristic approach to maximize modularity \cite{newman2006modularity} to discover the community structure. It can discover an appropriate community for each node towards the direction of increasing modularity. In modularity-based algorithms, Hollocou et al. propose a modularity-based sparse soft graph clustering algorithm \cite{hollocou2019modularity}. It can discover the community structure by maximizing the modularity of the nodes' partition and uses sparse matrix to record the partition result to improve the efficiency. Wang et al. propose a merging search algorithm based on modularity dominated density \cite{wang2015modularity}. Each node is partitioned into the community with the maximum modularity. After all nodes are divided into communities, a community revision is conducted. Multiple small communities are merged into a large community if the merger can improve the modularity of the community structure. A large community can also be split into multiple small communities if the split can improve the modularity. However, the community discovery algorithms based on modularity optimization only uses the neighborhood relationship of nodes and ignores the high-order information of a graph which can provide more semantic information. In addition, it is not easy to solve the optimization problem of the modularity. Therefore the above disadvantages restrict the applications of the modularity optimization in community discovery.

Matrix factorization is to decompose adjacency matrix or an information matrix related to adjacency matrix into multiple matrices. The community partition of the nodes can be directly obtained from the decomposition result. Therefore matrix factorization has a good explanation for the result of community discovery. For matrix factorization algorithms, Pei et al. propose a nonnegative matrix tri-factorization algorithm \cite{pei2015nonnegative}. It defines three interactive information matrices. Each interactive information matrix is decomposed into three matrices and introduces graph regularization to improve the performance of the algorithm. The partition result of the nodes is the fusion of the decomposition results of the three interactive information matrices. However, the time complexity of matrix factorization is high and it is not fit for a large scale network.

Graph neural network is a hot topic in recent years \cite{wu2020comprehensive,zhang2019graph}. It can obtain the low dimensional embedding vectors of nodes from the neighbors by one or multiple hidden layers. Due to the strong representation ability of neural network, graph neural network can effectively fuse the attribute and the structure information of a graph. In graph neural network algorithms, He et al. propose a community-centric graph convolutional network for community discovery \cite{zhang2020community}. It utilizes conditional random field model to design autocoder and introduces a local enhancement approach to make the similar nodes in the same community. Sun et al. propose a deep autoencoder-like nonnegative matrix factorization algorithm for community discovery \cite{ye2018deep}. It uses non-negative matrix factorization as the autoencoder and stacks multiple autoencoders to form the hierarchical decomposition result. However, in a graph, there is no tight connection between each node and the most other nodes and it means the adjacency matrix of a graph is high dimensional and sparse. If adjacency matrix is only as the initial input of graph convolution neural network \cite{kipf2017semi} without attribute information which is called as non-attributed graph, the sparse adjacency matrix cannot provide enough semantic information for graph convolution neural network.

In order to solve the restruction of the current algorithms and better discover the community structure, a deep graph convolution neural network with non-negative matrix factorization (DGCN-NMF) is proposed in this paper. DGCN-NMF employs multiple graph convolution layers to obtain the low dimensional embedding vectors of nodes. For each graph convolution layer, the outputs of the previous layers are decomposed into the low dimensional and dense matrices. The decomposition results of the previous layers and the output of the current layer are concatenated as the inputs of the next layers. The community partition can be obtained by an end-end-end approach. The main contributions of this paper are as follows:
\begin{itemize}
  \item A deep graph convolution neural network with self-supervised learning mechanism is proposed. DGCN-NMF can be trained from unlabeled nodes in a graph.
  \item Considering the concatenator increases the number of the parameters in a graph convolution neural network which can result in over-fitting, the output of each layer is decomposed into the low dimensional and dense matrices. The decomposition result is as one of the inputs of the next layers.
  \item The community assignment of each node can be outputted by an end-end-approach. The final community structure can be obtained by softmax approach.
\end{itemize}

The rests of this paper are organized as follows: Section 2 describes the detail theories and the steps of the proposed algorithm; Section 3 presents the experimental results and analyses. Section 4 concludes this paper and gives some research directions in the future.

\section{The proposed algorithm}
\begin{figure*}
  \centering
  % Requires \usepackage{graphicx}
  \includegraphics[width=18cm]{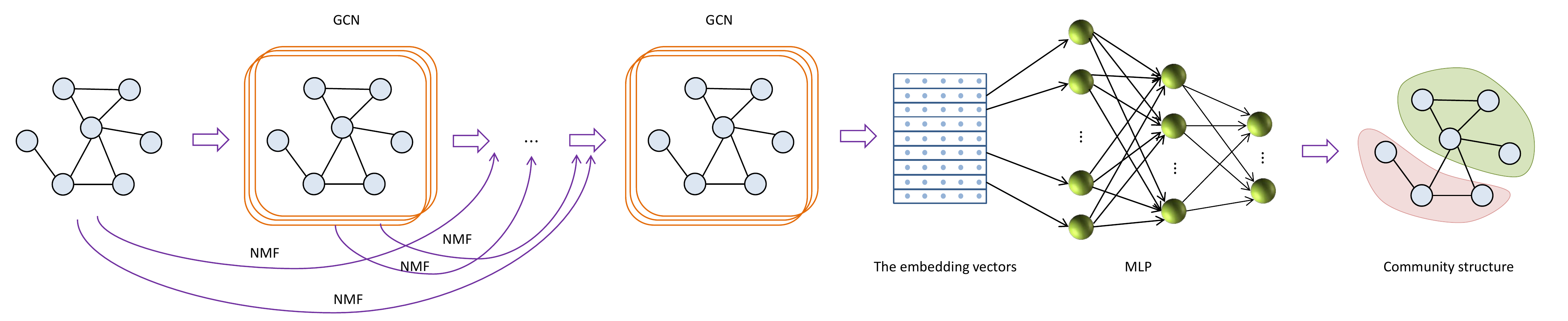}\\
  \caption{The structure of DGCN-NMF}\label{fg1}
\end{figure*}

Let $G=\left \langle V, E, \bm{A},\bm{X} \right \rangle$ be a graph. $V\in \mathbb{R}^{n}$ is the nodes of the graph \emph{G} and \emph{n} is the number of the nodes. \emph{E} is the edges of \emph{G}. $\bm{A}\in\mathbb{R}^{n\times n}$ is the adjacency matrix of \emph{G}. For $\forall v_{i}, v_{j}\in V$, $A_{ij}=1$ if $e_{ij}\in E$, otherwise, $A_{ij}=0$. $\bm{X}\in \mathbb{R}^{n\times m}$ is the attribute information of the nodes and \emph{m} is the dimensionality of the attribute information. The target of DGCN-NMF aims at learning a mapping function $f: V\mapsto \mathbb{R}^{n\times k}$. The nodes are assigned to the \emph{k} communities. The structure of DGCN-NMF is showed as Fig.\ref{fg1}.
\subsection{The graph convolution of DGCN-NMF}
Graph convolution neural network (GCN) can learn the low dimensional embedding vector of each node from the neighbors by the convolution and aggregation operators \cite{defferrard2016convolutional}. DGCN-NMF introduces multiple convolution layers to obtain the low dimensional embedding vectors of nodes. The convolution operator of each layer is as follow:
\begin{equation}\label{eq1}
  \bm{H}_{\ell+1}=\delta \left ( \widehat{\bm{D}}^{-\frac{1}{2}} \widehat{\bm{A}} \widehat{\bm{D}}^{-\frac{1}{2}}\bm{H}_{\ell}\bm{W}_{\ell}\right )
\end{equation}
where $\bm{H}_{\ell}$ is the output of the $\ell$th layer and it is also the input of the $(\ell+1)$th layer. $\widehat{\bm{A}}=\bm{A}+\bm{I}$ and $\bm{I}$ is an identity matrix. $\widehat{\bm{D}}$ is a diagonal matrix and $\widehat{D}_{ii}=\sum_{j=1}^{n}\widehat{A}_{ij}$. $\bm{W}_{\ell}$ is the parameters of the $\ell$ layer. $\bm{H}_{0}=\bm{X}$ if the graph \emph{G} is an attributed graph.  $\bm{H}_{0}=\bm{I}$ if the graph \emph{G} is a non-attributed graph. $\delta \left ( \cdot  \right )$ is an activation function.

It is known that $\bm{H}_{\ell+1}$ contains more high-level semantic information than $\bm{H}_{\ell}$. If $\bm{H}_{\ell+1}$ is as the output of DGCN-NMF, it means that only high-level semantic information is used and much low-level semantic information is ignored. For a graph, the low-level semantic information also plays an important role on the final embedding since the low-level semantic information can reflect the local structure information of a graph. Therefore both $\bm{H}_{\ell}$ and the outputs of the previous $(\ell-1)$ layers are as the input of the $(\ell+1)$th layer. However, the inputted dimension of the $(\ell+1)$ layer is large if the $\ell$ outputs are concatenated. The high dimension results in more parameters in graph convolution neural network and the over-fitting also appears.

In DGCN-NMF algorithm, the outputs of the previous $(\ell-1)$ layers are not concatenated directly. For $\forall \bm{H}_{i}$ $\left ( i=0,1,2,\cdots ,\ell-1 \right )$, it is decomposed into two matrices by non-negative matrix factorization (NMF) \cite{lee1999learning} as follow:
\begin{equation}\label{eq2}
  \bm{H}_{i} = \bm{U}_{i}\bm{V}_{i} \ \ \ s.t. \ \bm{U}_{i}\geq 0, \bm{V}_{i}\geq 0
\end{equation}
where $\bm{H}_{i}\in \mathbb{R}^{n\times L_{i}}$, $\bm{U}_{i}\in \mathbb{R}^{n\times d}$, $\bm{V}_{i}\in \mathbb{R}^{d\times L_{i}}$, $L_{i}$ is the outputted dimension of the \emph{i}th layer and \emph{d} is the output dimension of the non-negative matrix factorization. $\bm{U}_{i}$ is the low dimensional representation of $\bm{H}_{i}$ and $d\ll L_{i}$. Therefore the final the input of the $\left ( \ell+1 \right )$ is as:
\begin{equation}\label{eq3}
  \bm{H}_{\ell}\leftarrow \bm{H}_{\ell}\left |  \right | \bm{U}_{0}\left |  \right |\cdots\left |  \right |\bm{U}_{\ell-1}
\end{equation}
where $\left |  \right |$ is the concatenated operator and it can concatenate two matrices.

The low dimensional embedding vectors of nodes are outputted by GCNs. If the graph \emph{G} is an attributed graph, $\bm{X}$ is also concatenated to the final output of GCNs. Then the low dimensional embedding vectors are inputted into a MLP with single hidden layer. Let $\bm{Z}\in \mathbb{R}^{n\times k}$ be the output of MLP where \emph{k} is the number of communities. The community assignment of nodes can be determined from $\bm{Z}$ by \emph{softmax}.

\subsection{The self-supervised learning of DGCN-NMF}
In DGCN-NMF, the GCNs and MLP are trained from the unlabeled nodes of the graph \emph{G}. Self-supervised learning is introduced in this paper. The output $\bm{Z}$ of DGCN-NMF contains both the high-level and the low-level semantic information of the graph \emph{G}. Therefore the adjacency matrix of the graph \emph{G} should be restructured from $\bm{Z}$. The loss of the restruction $\mathcal{L}_{s}$ is defined as:
\begin{equation}\label{eq4}
  \mathcal{L}_{s} = \left \|\bm{Z}\bm{Z}^{T}-\bm{A} \right \|_{F}^{2}
\end{equation}
If the graph \emph{G} is an attributed graph, the loss of the restruction $\mathcal{L}_{s}$ is defined as:
\begin{equation}\label{eq5}
  \mathcal{L}_{s} = \left \| \bm{Z}\bm{Z}^{T} -\bm{A}\right \|_{F}^{2}+\left \| \bm{Z}\bm{Z}^{T}-\bm{X}\bm{X}^{T} \right \|_{F}^{2}
\end{equation}
In Eq.(\ref{eq5}), the first term is the restruction loss of the adjacency matrix. The second term is the restruction loss of the attribute information. DGCN-NMF is to restructure the structure and attribute information and minimize the restruction loss $\mathcal{L}_{s}$.

It is known that modularity is a measure for the strength of community structure. A large modularity means a better community structure. DGCN-NMF is to make the modularity of the community partition be as large as possible. The modularity $\mathcal{L}_{m}$ is as:
\begin{equation}\label{eq6}
  \mathcal{L}_{m} = \frac{1}{2e}\sum_{i,j}\left [ A_{ij}-\frac{k_{i}\cdot k_{j}}{2e} \right ]\sigma \left ( v_{i},v_{j} \right )
\end{equation}
where \emph{e} is the number of edges, $k_{i},k_{j}$ are the degrees of the nodes $v_{i},v_{j}$, respectively. $\sigma \left ( v_{i},v_{j} \right )=1$ if $v_{i}$ and $v_{j}$ are in the same community, otherwise, $\sigma \left ( v_{i},v_{j} \right )=0$.

Therefore DGCN-NMF is to discover a community partition with the minimum loss. The optimization problem of DGCN-NMF is as follow:
\begin{equation}\label{eq7}
  \underset{\bm{Z}}{\min}\ \mathcal{L}=\mathcal{L}_{s}-\alpha \mathcal{L}_{m}
\end{equation}
where $\alpha$ is a parameter. After the loss in Eq.(\ref{eq7}) is computed, the parameters of the GCNs and MLP in DGCN-NMF are adjusted and optimized by the back propagation approach.

Eqs.(\ref{eq1})-(\ref{eq7}) show the main steps of DGCN-NMF. It is known that the conventional GCN only outputs the high-level semantics features of a graph and DGCN-NMF can fuse both the high-level and the low-level semantics features. Therefore DGCN-NMF can use more structure information of a graph. %In addition, DGCN-NMF preserves the semantics feature and decreases the dimension to avoid over-fitting effectively.

%In Eq.(\ref{eq3}), NMF is introduced instead of Locally Linear Embedding (LLE), etc. The reason is that NMF is a linear dimension reduction approach and the different levels semantics can be well preserved.
\subsection{Time complexity analysis}
For DGCN-NMF, it costs $\mathcal{O}\left ( n^{3} \right )$ to compute $\widehat{\bm{D}}^{-\frac{1}{2}} \widehat{\bm{A}} \widehat{\bm{D}}^{-\frac{1}{2}}$. For each layer, it costs $\mathcal{O}\left ( nL_{\ell}L_{\ell-1}\right )$ to realize the nonlinear transformation. The time complexity of NMF is $\mathcal{O}\left ( ndL_{\ell}\right )$ for the $\ell$ layer. It costs $\mathcal{O}\left ( n^{2}\right )$ to compute Eqs.(\ref{eq4})-(\ref{eq6}). Let $d^{'}$ be the input dimension of MLP. The time complexity of MLP is $\mathcal{O}\left ( nd^{'}k\right )$. Therefore the time complexity of DGCN-NMF is $\mathcal{O}\left ( n^{3}t\right )$ where \emph{t} is the total epoches.
\section{The experimental results and analyses}
In order to test the performance of DGCN-NMF, STCD \cite{bai2018novel}, MDDBMS \cite{wang2015modularity}, MNMF \cite{wang2017community} and ESCG \cite{liu2013large} are selected as the comparison algorithms. For DGCN-NMF, it uses double hidden layers and the outputted dimensions are as $\left [ 150, 100 \right ]$. The activation function of MLP is \emph{sigmoid}. $d=64$ and $\alpha\in\left [ 0,100 \right ]$. DGCN-NMF-s is the DGCN-NMF algorithm with \emph{sigmoid} activation function for GCNs. DGCN-NMF-r is the DGCN-NMF algorithm with \emph{ReLU} activation function for GCNs. Adjnoun, football and politics are selected as the experimental data sets. Owing to the restriction of the layout, the descriptions of the data sets are not showed in this paper and the detailed descriptions can be seen from the websites\footnote{\url{http://memetracker.org/data/index.html}}$^{,}$\footnote{\url{http://www-personal.umich.edu/\%7Emejn/netdata/}}.

In this paper, Jaccard index (J), Folkes and Mallows index (FM), Kulczynski index (K), recall and F1-measure (F1) are as the evaluation criteria. The descriptions of the evaluation criteria can be seen from the reference \cite{Xie2016Unsupervised}.

\begin{table}[h]
\small
\centering
\caption{The experimental results of the algorithms on adjnoun data set.}
\begin{tabular}{ccccccccccccc}\toprule
	&J	&FM	&K	&recall	&F1	\\\midrule
STCD	&0.0135	&0.1292	&0.3202		&0.0272	&0.0522	\\
MDDBMS	&0.0145	&0.1208	&0.2646 &0.0292	&0.0551	\\
MNMF	&0.2458	&0.4942	&0.4942	&0.4953	&0.4942	\\
ESCG	&0.3953	&0.6271	&0.6463	&0.7967	&0.6086	\\\hline
DGCN-NMF-r	&\textbf{0.4781}	&\textbf{0.6911}	&\textbf{0.7297}	&\textbf{0.9637}	&\textbf{0.6546}	\\
DGCN-NMF-s	&0.2476	&0.4954	&0.4954 &0.4990	&0.4954	\\\toprule
\end{tabular}\label{tb1}
\end{table}
\begin{table}[h]
\small
\centering
\caption{The experimental results of the algorithms on football data set.}
\begin{tabular}{ccccccccccccc}\toprule
	&J	&FM	&K	&recall	&F1	\\\midrule
STCD	&0.0188	&0.4054	&0.4670		&0.2352	&0.3519 \\
MDDBMS	&0.0279	&0.4494	&0.4636	&0.3499	&0.4357	\\
MNMF	&0.0571	&0.4487	&0.4983		&0.7151	&0.4040	\\
ESCG	&0.0239	&0.1541	&0.1895		&0.2994	&0.1255	\\\hline
DGCN-NMF-r	&\textbf{0.0581}	&0.5397	&0.5641	&\textbf{0.7285}	&0.5163\\
DGCN-NMF-s	&0.0572	&\textbf{0.5611}	&\textbf{0.5781}	&0.7170	&\textbf{0.5447}\\\toprule
\end{tabular}\label{tb2}
\end{table}
\begin{table}[h]
\small
\centering
\caption{The experimental results of the algorithms on politics data set.}
\begin{tabular}{ccccccccccccc}\toprule
	&J	&FM	&K	&recall	&F1	\\\midrule
STCD	&0.0110	&0.1453	&0.3937		&0.0278	&0.0537	\\
MDDBMS	&0.0441	&0.3074	&0.4787	&0.1117	&0.1974	\\
MNMF	&0.2673	&0.7044	&0.7052		&0.6767	&0.7037	\\
ESCG	&0.2529	&0.5053	&0.5214		&0.6401	&0.4900	\\\hline
DGCN-NMF-r	&\textbf{0.3273}	&0.7897	&0.7906	&\textbf{0.8285}	&0.7888\\
DGCN-NMF-s	&0.3038	&\textbf{0.7960}	&\textbf{0.7964}	&0.7691	&\textbf{0.7955}\\\toprule
\end{tabular}\label{tb3}
\end{table}

Tables \ref{tb1}-\ref{tb3} shows the experimental results. From the results, it is known that DGCN-NMF outperforms the comparison algorithms on the experimental data sets and the advantages of DGCN-NMF are obvious. The experimental results demonstrate that DGCN is an effective approach for community discovery. MNMF is a community discovery algorithm based on non-negative matrix factorization. It can be seen that DGCN-NMF outperforms MNMF. It means that the approach that introduces deep graph convolution neural network into NMF can improve the performance of the algorithm. MDDBMS is a representative community discovery algorithm based on modularity optimization. On the three experimental data sets, the performance of DGCN-NMF is much better than MDDBMS. The results prove that it cannnot find community structure well only relying on the low-level semantic information of a graph. DGCN-NMF can fuse both the low-level and the high-level semantic information. Therefore it can obtain the best results on the experimental data sets.

By analyzing the results of DGCN-NMF-r and DGCN-NMF-s, it can be seen that the activation function plays an important role on the performance of DGCN-NMF. The performance of DGCN-NMF changes with different activation functions. Therefore it is of great significance for DGCN-NMF to select an appropriate activation function.
\section{Conclusions}
In this paper, a deep graph convolution neural network with non-negative matrix factorization (DGCN-NMF) for community discovery is proposed. DGCN-NM employs multiple graph convolution layers to obtain the low dimensional embedding vectors of nodes. In addition, the output matrix of each hidden layer is decomposed by NMF and the decomposition result is as one of the inputs for the next layers. Therefore DGCN-NM can use bot the low level and the high level semantics information effectively. The final community partition is outputted by an end-to-end approach. The experimental results demonstrate that DGCN-NMF is an effective approach for community discovery.

Pre-training is an effective approach to improve the performance of graph neural network. In the future, we will introduce pre-training into DGCN-NMF. Multi-view learning is to learn knowledge and patterns from multiple views. Therefore how to extend DGCN-NMF to multi-view learning is also a potential research issue.
\section{Acknowledgments}
This work was supported by National Natural Science Fund of China (Nos. 61972064, 61672130), Liaoning Revitalization Talents Program (No. XLY-C1806006), Fundamental Research Funds for the Central Universities (No. DUT19RC(3)012).

\bibliographystyle{IEEEbib}
\bibliography{ref}
\end{document}